\documentclass[12pt]{article}

\usepackage{amssymb}
\usepackage{amsthm}
\usepackage{amsmath} %

\usepackage{bm,stmaryrd}% bold math
\usepackage[linkcolor=blue,colorlinks=true]{hyperref}
\usepackage{hyperref}% add hypertext capabilities

\usepackage{tikz}
\usepackage{pgfplots}
\usepackage{float}
\usetikzlibrary{plotmarks}
\usetikzlibrary{patterns}
\usetikzlibrary{arrows}

\newcommand{\bsigma}{\bm \sigma}
\newcommand{\cave}{\la c \ra}
\newcommand{\cgr}{c_\text{g}}
\newcommand{\cph}{c_\text{ph}}
\newcommand{\cm}{{c_{-}}}
\newcommand{\cp}{{c_{+}}}
\newcommand{\de}{\partial}
\newcommand{\dst}{\displaystyle}
\newcommand{\ep}{\varepsilon}
\newcommand{\epp}{\ep_{+}}
\newcommand{\epm}{\ep_{-}}
\newcommand{\mup}{\mu_{+}}
\newcommand{\mum}{\mu_{-}}
\newcommand{\n}{\bm n}

\newcommand{\om}{\omega}

\newcommand{\kp}{{k_{+}}}
\newcommand{\oh}{\frac{1}{2}}
\newcommand{\la}{\left \langle}
\newcommand{\ra}{\right \rangle}
\newcommand{\rf}[1]{(\ref{#1})}
\newcommand{\tp}{t_{+}}
\newcommand{\tm}{t_{-}}
\renewcommand{\u}{\bm u}
\newcommand{\um}{u^{-}}
\newcommand{\up}{u^{+}}
\newcommand{\vr}{\varrho}
\renewcommand{\rho}{\vr}
\newcommand{\vre}{\vr_{+}}
\newcommand{\vri}{\vr_{-}}
\newcommand{\z}{\bm 0}
\newcommand{\zp}{z_{+}}
\newcommand{\zm}{z_{-}}
\renewcommand{\leq}{\leqslant}

\newcommand{\I}{\ensuremath{\mathrm{i}}}
\newcommand{\E}{\ensuremath{\mathrm{e}}}
\newcommand{\D}{\ensuremath{\mathrm{d}}}

\begin{document}

%%%% Article title to be placed here
\title{Slowdown of group velocity in periodic waveguides\\}
\author{Yuri A. Godin\thanks{Department of Mathematics and Statistics, University of North Carolina at Charlotte, 
Charlotte, NC 28223 USA. E-mail: ygodin@uncc.edu}~ and Boris Vainberg\thanks{Department of Mathematics and Statistics, 
University of North Carolina at Charlotte, 
Charlotte, NC 28223 USA. Email: brvainbe@uncc.edu}}

\date{\today}
\maketitle

% \date{\today}% It is always \today, today, April 27, 2020

\begin{abstract}
We consider the propagation of acoustic, electromagnetic and elastic waves in a one-dimensional periodic two-component material. Accurate asymptotic formulas are provided for the
group velocity as a function of the material parameters when the concentration of scatterers is small
or the characteristic impedances of the two media differ substantially. In the latter case, it is shown that the minimum group velocity occurs when the volume fractions of the components of the material are equal. 
In both asymptotic cases we show that the leading terms of the group velocity do not depend on frequency.
Thus slowdown is frequency-independent and is not related to the resonance phenomena.

\end{abstract}


\maketitle

\section{Introductions}

Periodic 
media offer a great deal of possibilities for manipulating wave propagation. The spectrum of propagating
frequencies consists of closed intervals (bands) and the propagation of waves is suppressed for the frequencies in complimentary intervals (gaps).
While propagating waves do not exist in the frequency gaps, the character of waves in the bands can be essentially different
in terms of their direction, speed, and amplitude of propagation. One example of such a highly dispersive medium
represents a fluid containing a periodic lattice of scatterers. The propagation of acoustic waves in such media through a lattice
of air bubbles or cylinders of air has been studied numerically \cite{Ruffa:92,Kafesaki:00,Krokhin:2003,Torrent:2006,Torrent:2007,Leroy:2009,Skvortsov:2019},
along with some analytical approaches \cite{Belyaev:87,Ye:03,Torrent:2012,Boutin:2013} and more complicated geometries \cite{Humphrey:1998,Wu:02,Bennetts:2019}.

A related one-dimensional periodic model has been studied by the well-known transfer (or propagation) matrix formalism \cite{Hale:80} for acoustic materials \cite{Ye:00, Vasseur:01, Khelif:04, Olsson:09},  for periodic elastic media \cite{Cocoletzi:94,Sigalas:95,Hussein:06,Boudouti:09}, and for photonic band-gap structures \cite{Bendickson:96, Yeh:05}. More complicated problems were considered in \cite{Bradley:94, Adams:08}. Dispersion relations obtained in most of the papers above were studied numerically. 
Our main goal is to study the dispersion relation analytically and  provide asymptotic formulas for the group velocity when the concentration of scatterers is small or the ratio of the impedances is large. 
The main terms in both asymptotic formulas do not depend on the frequency. Thus, a slowdown occurs at any frequency in any band and does not have a resonance character. We show that the smallest group velocity is achieved when the scatterers and the host material are taken in equal proportions regardless of their physical properties provided that the impedance ratio is large. All the asymptotic formulas are verified numerically.

\section{Acoustic waves}

\begin{figure}[H]
\begin{center}
  \begin{tikzpicture}[x={(-10:1cm)},y={(90:1cm)},z={(210:1cm)},>=latex]
    % Axes
    \draw (2.5,0,0) node[above] {$x$} -- (5,0,0);
    \draw (4,0,0) -- (4,2,0) node[above] {$z$};
    \draw (4,0,0) -- (4,0,2) node[left] {$y$};
    \foreach \i in {0,1,...,3}{
    \draw[fill=white!40!black] (5+\i,0,2) -- (5.5+\i,0,2)--(5.5+\i,2,2)--(5+\i,2,2)--cycle;
    \draw[fill=white!40!black] (5.5+\i,2,2) -- (5.5+\i,2,0.5)--(5+\i,2,0.5)--(5+\i,2,2)--cycle;
    \draw[fill=white!80!black] (5.5+\i,0,2) -- (6+\i,0,2)--(6+\i,2,2)--(5.5+\i,2,2)--cycle;
    \draw[fill=white!80!black] (6+\i,2,2) -- (6+\i,2,0.5)--(5.5+\i,2,0.5)--(5.5+\i,2,2)--cycle;
    }
    \draw[fill=white!80!black] (9,0,2) -- (9,0,0.5)--(9,2,0.5)--(9,2,2)--cycle;
    % Propagation
    \draw[->,ultra thick] (8,0,0) -- node[above] {$c$} (9,0,0);
    \draw[<->] (5,0,2.5) -- (6,0,2.5) node[midway,below]{$\ell$};
  \end{tikzpicture}
\end{center}
\caption{One-dimensional periodic medium with alternating layers.}
\label{PC}
\end{figure}
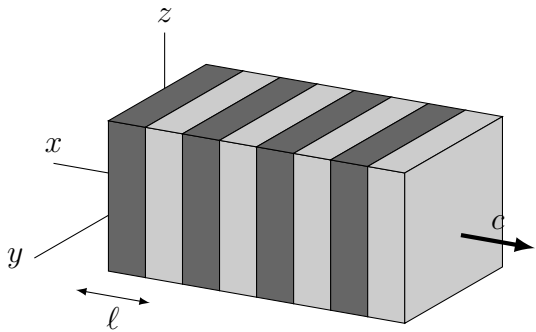

We consider the propagation of acoustic waves through a one-dimensional periodic medium with period $\ell$ consisting of two materials
with mass densities $\vr_{\mp}$ and wave speed $c_{\mp}$ when $n\ell  <x <n\ell + \delta$ and $n\ell + \delta < x < (n+1)\ell$, $n \in {\mathbb Z}$, respectively (see Fig. \ref{PC}).

Assuming that the excess pressure $p(x, t)$ is time-harmonic $\dst p(x, t) = u(x) \E^{-\I \omega t},$
the amplitude $u(x)$ of acoustic wave satisfies the equation
\begin{align}
\label{eq1}
 \frac{\D^2 \um}{\D\, x^2} + k^2_{-} \um &= 0, \quad n\ell  <x <n\ell + \delta, \\[2mm]
 \frac{\D^2 \up}{\D\, x^2} + k^2_{+} \up &= 0, \quad n\ell + \delta < x < (n+1)\ell,
\label{eq2}
\end{align}
subject to the boundary conditions across all the interfaces $\{n\ell\}$ and $\{n\ell + \delta\}$:
\begin{align}
\label{bc}
\um = \up, \quad \frac{1}{\vr_{+}}\, \up_x  = \frac{1}{\vr_{-}}\, \um_x,
\end{align}
where $k_{\pm} = \dfrac{\om}{c_{\pm}}$. %$c_{\pm}$ is the wave speed and $\vr_{\pm}$ is the mass density of the corresponding medium.
We are looking for the Floquet solutions of \rf{eq1}-\rf{bc} satisfying the condition
\begin{align}
\label{eq4}
  \E^{\I\kappa x} u ~~{\text{is periodic with the period}}~~ \ell,
\end{align}
with some real  $\kappa$ (the wavenumber).

We will assume that  $\ell =1$.
The eigenvalues $\lambda$ of the transfer matrix over the period satisfy the characteristic equation
\begin{align}
\lambda^2 -2\lambda F + 1=0,
 \label{char_eqn}
\end{align}
where
\begin{align}
F &= \cos \frac{(1-\delta)\,\om}{\cp} \cos \frac{\om \delta}{\cm} 
-\oh \left(\frac{\zp}{\zm} + \frac{\zm}{\zp}\right)
 \sin \frac{(1-\delta)\,\om}{\cp} \sin \frac{\om \delta}{\cm}
\end{align}
and
 \begin{align}
 z_{\pm} &= c_{\pm} \vr_{\pm}
\end{align}
are characteristic acoustic impedances of the two media, respectively.

Solution of \rf{char_eqn} in the form $\lambda = \E^{\I \kappa}$ with real $\kappa$ leads to the relation
\begin{align}
 \label{F}
 \cos \kappa = F.
\end{align}
The inequality $|F| \leq 1$ provides the necessary and sufficient condition for which problem \rf{eq1}-\rf{eq4} has a nontrivial solution. The intervals of $\om$ where this condition holds are called the bands. 
From \rf{F} we can find the group velocity $\cgr = \dfrac{\de \om}{\de \kappa}$ in any band:
\begin{align}
 \cgr &= -\frac{\sin \kappa}{\frac{\de F}{\de \om}}.
 \label{cgr}
\end{align}
Denoting
\begin{align}
 \tp &= \frac{\delta}{\cm} + \frac{1-\delta}{\cp}, \quad \tm = \frac{1-\delta}{\cp} -\frac{\delta}{\cm}, \\[2mm]
 \eta &= \frac{1}{2}\left(\frac{\zp}{\zm} + \frac{\zm}{\zp}\right),
 \label{sigma}
\end{align}
the group velocity \rf{cgr} takes the form
\begin{align}
 \cgr = 2\, \frac{\sqrt{1-\frac{1}{4}\left((1+\eta)\cos \om \tp + (1-\eta)\cos \om \tm \right)^2}}{(1+\eta)\,\tp \sin \om \tp + (1-\eta)\,\tm \sin \om \tm}.
 \label{Cgr}
\end{align}

\subsection*{Equal impedances}

If the characteristic impedances of the two media are equal, $\cp \vre = \cm \vri$, then formula \rf{Cgr} gives $\cgr = t_{+}^{-1}$. This means that the group velocity coincides with the velocity of a wave that propagates through the interval $(0,\delta)$ with the speed $\cm$ and over the interval $(\delta,1)$ with the speed $\cp$ without slowing down or acceleration. From \rf{char_eqn} we have
\begin{align}
 F =  \cos \left(\frac{\delta}{\cm} + \frac{1-\delta}{\cp} \right) \om
\end{align}
and
\begin{align}
 \kappa = \left(\frac{\delta}{\cm} + \frac{1-\delta}{\cp}  \right) \om.
\end{align}
Then the group velocity $\cgr$, the phase velocity $\cph = \dfrac{\om}{\kappa}$, and the average speed of propagation $\cave$ are all coincide and are equal to
\begin{align}
 \cave = \left(\frac{\delta}{\cm} + \frac{1-\delta}{\cp}  \right)^{-1}.
 \label{cave}
\end{align}
Here, waves propagate without reflection and the speed does not depend on the wave frequency.

\subsection*{Small scatterers}

Let us find an approximation of the eigenvalues $\lambda$ of the transfer matrix when $\delta \ll 1$.
We approximate \rf{char_eqn} in any fixed band by
\begin{align}
 \lambda^2 -\lambda
 \left(2 \cos \kp -\left( \frac{\zp}{\zm} + \frac{\zm}{\zp}\right) \frac{\om \delta}{\cm}\, \sin \kp  \right) + 1 = 0.
\end{align}
Then
\begin{align}
  \lambda = \E^{\I \kappa} = \E^{\I \kp} \left( 1 + \frac{\I}{2}
  \left( \frac{\zp}{\zm} + \frac{\zm}{\zp}\right) \frac{\om \delta}{\cm} \right) + O\left( \delta^2 \right).
\end{align}
and
\begin{align}
 \kappa = \frac{\om}{\cp} \left(1+\frac{1}{2}\left( \frac{\zp}{\zm} + \frac{\zm}{\zp}\right) \frac{\cp}{\cm} \,\delta + O\left(\delta^2 \right)\right).
 \label{kappa_soft}
\end{align}
Expression \rf{kappa_soft} allows us to find the group velocity $\cgr$ (which coincides with the phase velocity in this approximation)
\begin{align}
 \cgr = \cp  \left(1-\frac{1}{2}\, \frac{\cp}{\cm} \left( \frac{\zp}{\zm} + \frac{\zm}{\zp}\right) \delta + O\left(\delta^2 \right)\right), \quad \delta \ll 1.
 \label{cs}
\end{align}

\begin{figure}[ht]
\centering
\begin{tikzpicture}%[
\begin{axis}[%
xmin = 0,
xmax = 0.00005,
xtick={0, 1.0e-05, 2.0e-05, 3.0e-05, 4.0e-05, 5.0e-05},
ymin = 0.6,
ymax = 1,
xlabel=$\delta$,
ylabel=$\cgr/\cp$,
grid=major,
]
\addplot+[blue,line width= 2pt, dashed, dash pattern=on 5pt off 5pt,no marks,domain=0.0:0.0001] {1-6521.1605*x};
\addplot [color=red,solid,line width= 2pt, forget plot,smooth] plot coordinates{
(0.00, 1)
(1.00000e-05, 0.940553)
(2.00000e-05, 0.890588)
(3.00000e-05, 0.847825)
(4.00000e-05, 0.810682)
(5.00000e-05, 0.778043)
(6.00000e-05, 0.749032)
(7.00000e-05, 0.723035)
(8.00000e-05, 0.699599)
(9.00000e-05, 0.678292)
(0.0001, 0.65882)
};
\end{axis}%
\end{tikzpicture}
\caption{Dependence of the normalized group velocity of a one-dimensional periodic acoustic waveguide
consisting of air layers in water as a function of the volume fraction $\delta$ of air.
The solid red line corresponds to exact velocity \rf{Cgr}, while the blue dashed line corresponds to the linear approximation \rf{cs}.}
\label{graph1}
\end{figure}
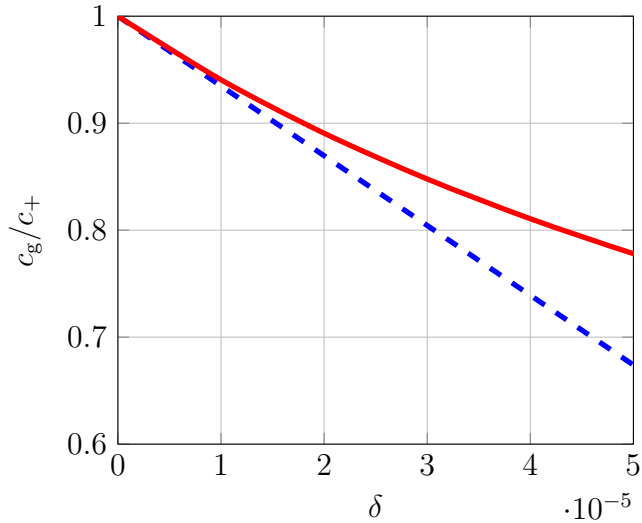

The coefficient in front of $\delta$ in \rf{cs} can be huge in magnitude for soft scatterers that results in a dramatic
group velocity reduction even for a tiny concentration of the scatterers. Such an example is shown in Figure \ref{graph1}
where an acoustic wave with frequency $\om = 10\,$s$^{-1}$ propagates through a periodic waveguide consisting of layers of
the air ($\vri = 1.3\,$kg/m$^3$, $\cm = 340\,$m/s) of width $\delta$ followed by a layer of water
($\vre = 1000\,$kg/m$^3$, $\cp = 1400\,$m/s). The solid red line shows the exact value of the group velocity \rf{Cgr}
while the blue dashed line corresponds to the linear approximation by \rf{cs} for very small concentration of the air.
A similar result is valid for two and three-dimensional periodic media
\cite{GV:2019} as well as for the random waveguide \cite{YG:06}.

\subsection*{Highly mismatched impedances}

When the characteristic impedances of the two media differ substantially, i.e. $\eta \gg 1$, and in addition 
$\gamma_{\pm} = \dfrac{\eta \,\om^2}{c^2_{\pm}} \ll 1$, 
the group velocity away from the points $\delta = 0$ and $\delta = 1$ is approximated by
\begin{align}
 \cgr = \sqrt{\frac{\cp \, \cm}{2\, \delta \,(1-\delta)\, \eta}}.
 \label{cgr_asy}
\end{align}
The latter formula agrees with (16) with an accuracy of $O\left(\eta^{-1} + \gamma_{+} + \gamma_{-}\right)$.
The minimum of \rf{cgr_asy} is attained at $\delta = \oh$ and does not depend on $\om$. Using \rf{sigma} we have
\begin{align}
 \min \cgr = 2\left(\frac{\cp \, \cm}{ \frac{\zp}{\zm} + \frac{\zm}{\zp}}\right)^{\oh}
 = 2\,\cp \,\cm \left(\frac{\vre \vri}{c^2_{+} \vre^2 + c^2_{-} \vri^2 }\right)^{\oh}.
\label{cgr_min}
\end{align}

\begin{figure}[ht]
\centering
\begin{tikzpicture}
\begin{axis}[%
xlabel=$\delta$,
ylabel=$\cgr/\cp$,
xtick={0, 0.1, 0.2, 0.3, 0.3, 0.4, 0.5, 0.6, 0.7, 0.8, 0.9, 1.0},
ytick={0, 0.1, 0.2, 0.3, 0.3, 0.4, 0.5, 0.6, 0.7, 0.8, 0.9, 1.0},
grid=major,
xmin=0, xmax=1.0,
ymin=0.0, ymax=1.0,
]
\addplot [black,line width= 1.5pt,domain=0:1] {1000/(2000*x + (1-x)*1000)}; % rigid

\addplot [draw=red, line width= 1pt, mark=*, mark options={fill=white}] plot coordinates{
(0.00, 1)
(0.01, 0.918963)
(0.02, 0.853574)
(0.03, 0.79935)
(0.04, 0.753426)
% (0.05, 0.713878)
(0.06, 0.679357)
% (0.07, 0.648887)
(0.08, 0.621739)
% (0.09, 0.597361)
% (0.1, 0.57532)
(0.11, 0.555277)
% (0.12, 0.536958)
% (0.13, 0.520142)
(0.14, 0.504645)
% (0.15, 0.490316)
% (0.16, 0.477028)
(0.17, 0.464673)
% (0.18, 0.45316)
% (0.19, 0.44241)
(0.2, 0.432356)
% (0.21, 0.422939)
% (0.22, 0.414107)
(0.23, 0.405817)
% (0.24, 0.398028)
% (0.25, 0.390706)
(0.26, 0.38382)
% (0.27, 0.377342)
% (0.28, 0.371249)
(0.29, 0.365519)
% (0.3, 0.360132)
% (0.31, 0.35507)
(0.32, 0.350319)
% (0.33, 0.345864)
% (0.34, 0.341692)
(0.35, 0.337792)
% (0.36, 0.334154)
% (0.37, 0.330767)
(0.38, 0.327625)
% (0.39, 0.324718)
% (0.4, 0.322041)
(0.41, 0.319586)
% (0.42, 0.317349)
% (0.43, 0.315324)
(0.44, 0.313507)
% (0.45, 0.311893)
% (0.46, 0.310479)
(0.47, 0.309261)
% (0.48, 0.308236)
% (0.49, 0.307402)
(0.5, 0.306756)
% (0.51, 0.306296)
% (0.52, 0.30602)
% (0.53, 0.305927)
(0.54, 0.306014)
% (0.55, 0.306282)
% (0.56, 0.306728)
% (0.57, 0.307352)
(0.58, 0.308152)
% (0.59, 0.30913)
% (0.6, 0.310283)
% (0.61, 0.311613)
(0.62, 0.313118)
% (0.63, 0.314798)
% (0.64, 0.316654)
% (0.65, 0.318686)
(0.66, 0.320894)
% (0.67, 0.323278)
% (0.68, 0.325839)
% (0.69, 0.328577)
(0.7, 0.331492)
% (0.71, 0.334586)
% (0.72, 0.337859)
% (0.73, 0.341312)
(0.74, 0.344945)
% (0.75, 0.348758)
% (0.76, 0.352754)
% (0.77, 0.356931)
(0.78, 0.36129)
% (0.79, 0.365833)
% (0.8, 0.370559)
% (0.81, 0.375468)
(0.82, 0.380559)
% (0.83, 0.385834)
% (0.84, 0.39129)
% (0.85, 0.396927)
(0.86, 0.402742)
% (0.87, 0.408735)
% (0.88, 0.414901)
% (0.89, 0.421238)
(0.9, 0.427741)
% (0.91, 0.434404)
% (0.92, 0.441222)
% (0.93, 0.448185)
(0.94, 0.455286)
% (0.95, 0.462512)
% (0.96, 0.469852)
(0.97, 0.477289)
% (0.98, 0.484807)
% (0.99, 0.492385)
(1, 0.5)
};

\addplot [draw=blue, line width= 2pt, dash pattern={on 7pt off 2pt on 1pt off 2pt}, no marks] plot coordinates{
(0.00, 1)
(0.01, 0.57727)
(0.02, 0.445309)
(0.03, 0.37461)
(0.04, 0.328802)
(0.05, 0.295993)
(0.06, 0.270984)
(0.07, 0.251092)
(0.08, 0.234774)
(0.09, 0.221072)
(0.1, 0.209353)
(0.11, 0.199183)
(0.12, 0.19025)
(0.13, 0.182326)
(0.14, 0.175238)
(0.15, 0.168853)
(0.16, 0.163067)
(0.17, 0.157795)
(0.18, 0.152972)
(0.19, 0.148542)
(0.2, 0.14446)
(0.21, 0.140687)
(0.22, 0.137193)
(0.23, 0.133951)
(0.24, 0.130936)
(0.25, 0.128131)
(0.26, 0.125517)
(0.27, 0.12308)
(0.28, 0.120808)
(0.29, 0.118688)
(0.3, 0.116712)
(0.31, 0.114871)
(0.32, 0.113156)
(0.33, 0.111563)
(0.34, 0.110084)
(0.35, 0.108715)
(0.36, 0.10745)
(0.37, 0.106287)
(0.38, 0.10522)
(0.39, 0.104248)
(0.4, 0.103367)
(0.41, 0.102575)
(0.42, 0.101869)
(0.43, 0.101249)
(0.44, 0.100712)
(0.45, 0.100256)
(0.46, 0.099882)
(0.47, 0.0995873)
(0.48, 0.0993718)
(0.49, 0.0992349)
(0.5, 0.0991762)
(0.51, 0.0991956)
(0.52, 0.099293)
(0.53, 0.0994686)
(0.54, 0.0997229)
(0.55, 0.100056)
(0.56, 0.100469)
(0.57, 0.100963)
(0.58, 0.101539)
(0.59, 0.102198)
(0.6, 0.102942)
(0.61, 0.103772)
(0.62, 0.104691)
(0.63, 0.105701)
(0.64, 0.106804)
(0.65, 0.108005)
(0.66, 0.109306)
(0.67, 0.110712)
(0.68, 0.112226)
(0.69, 0.113854)
(0.7, 0.115602)
(0.71, 0.117476)
(0.72, 0.119483)
(0.73, 0.121632)
(0.74, 0.123932)
(0.75, 0.126393)
(0.76, 0.129028)
(0.77, 0.131852)
(0.78, 0.134879)
(0.79, 0.138129)
(0.8, 0.141624)
(0.81, 0.145387)
(0.82, 0.149451)
(0.83, 0.153849)
(0.84, 0.158623)
(0.85, 0.163824)
(0.86, 0.169514)
(0.87, 0.175767)
(0.88, 0.182678)
(0.89, 0.190364)
(0.9, 0.198978)
(0.91, 0.208717)
(0.92, 0.219847)
(0.93, 0.232732)
(0.94, 0.247888)
(0.95, 0.266075)
(0.96, 0.28847)
(0.97, 0.317019)
(0.98, 0.355227)
(0.99, 0.410258)
(1, 0.5)
};

% \addplot+[draw=green, line width= 2pt,no marks, forget plot] plot coordinates{
\addplot [draw, color=black!30!green, line width= 2pt, dash pattern={on 6pt off 4pt on 6pt off 4pt}, no marks] plot coordinates{
% cp = 2000, `&varrho;p` = 1000, cm = 1000, `&varrho;m` = 2 omega = 130
(0.00, 1)
(0.01, 0.217168)
(0.02, 0.154569)
(0.03, 0.12603)
(0.04, 0.108779)
(0.05, 0.0968947)
(0.06, 0.0880584)
(0.07, 0.08115)
(0.08, 0.0755536)
(0.09, 0.0708985)
(0.1, 0.0669466)
(0.11, 0.0635369)
(0.12, 0.0605563)
(0.13, 0.0579224)
(0.14, 0.0555738)
(0.15, 0.0534636)
(0.16, 0.0515555)
(0.17, 0.0498204)
(0.18, 0.0482352)
(0.19, 0.0467811)
(0.2, 0.0454427)
(0.21, 0.0442069)
(0.22, 0.0430632)
(0.23, 0.0420025)
(0.24, 0.0410169)
(0.25, 0.0401)
(0.26, 0.0392461)
(0.27, 0.0384503)
(0.28, 0.0377083)
(0.29, 0.0370165)
(0.3, 0.0363717)
(0.31, 0.0357709)
(0.32, 0.0352119)
(0.33, 0.0346923)
(0.34, 0.0342104)
(0.35, 0.0337643)
(0.36, 0.0333526)
(0.37, 0.032974)
(0.38, 0.0326273)
(0.39, 0.0323116)
(0.4, 0.0320259)
(0.41, 0.0317695)
(0.42, 0.0315416)
(0.43, 0.0313418)
(0.44, 0.0311695)
(0.45, 0.0310243)
(0.46, 0.0309058)
(0.47, 0.0308138)
(0.48, 0.0307481)
(0.49, 0.0307085)
(0.5, 0.0306949)
(0.51, 0.0307072)
(0.52, 0.0307456)
(0.53, 0.0308101)
(0.54, 0.0309008)
(0.55, 0.031018)
(0.56, 0.0311619)
(0.57, 0.0313329)
(0.58, 0.0315313)
(0.59, 0.0317577)
(0.6, 0.0320126)
(0.61, 0.0322967)
(0.62, 0.0326107)
(0.63, 0.0329556)
(0.64, 0.0333323)
(0.65, 0.033742)
(0.66, 0.0341859)
(0.67, 0.0346656)
(0.68, 0.0351827)
(0.69, 0.035739)
(0.7, 0.0363367)
(0.71, 0.0369783)
(0.72, 0.0376665)
(0.73, 0.0384044)
(0.74, 0.0391958)
(0.75, 0.0400448)
(0.76, 0.0409561)
(0.77, 0.0419354)
(0.78, 0.0429891)
(0.79, 0.0441247)
(0.8, 0.0453511)
(0.81, 0.0466789)
(0.82, 0.0481205)
(0.83, 0.0496911)
(0.84, 0.051409)
(0.85, 0.0532967)
(0.86, 0.0553822)
(0.87, 0.0577008)
(0.88, 0.0602978)
(0.89, 0.0632322)
(0.9, 0.0665829)
(0.91, 0.0704578)
(0.92, 0.0750099)
(0.93, 0.0804635)
(0.94, 0.0871653)
(0.95, 0.0956845)
(0.96, 0.10704)
(0.97, 0.123295)
(0.98, 0.149508)
(0.99, 0.203514)
(1, 0.5)
};

% \addplot+[draw=magenta, line width= 2pt,no marks, forget plot] plot coordinates{
\addplot [draw=violet, line width= 2pt, dash pattern={on 2pt off 2pt on 2pt off 2pt}, no marks] plot coordinates{
% \addplot+[draw=violet, line width= 3pt, dash pattern={on 7pt off 7pt on 7pt off 7pt}, no marks] plot coordinates{
% cp = 2000, `&varrho;p` = 1000, cm = 1000, `&varrho;m` = 0.2 omega = 40
(0.00, 1)
(0.001, 0.218124)
(0.002, 0.156029)
(0.003, 0.127855)
(0.004, 0.110898)
(0.005, 0.0992632)
(0.006, 0.0906442)
(0.007, 0.0839281)
(0.008, 0.0785034)
(0.009, 0.0740026)
(0.01, 0.0701897)
(0.02, 0.0494495)
(0.03, 0.0401963)
(0.04, 0.0346517)
(0.05, 0.0308513)
(0.06, 0.0280351)
(0.07, 0.025839)
(0.08, 0.0240635)
(0.09, 0.0225892)
(0.1, 0.0213395)
(0.11, 0.0202628)
(0.12, 0.0193227)
(0.13, 0.0184929)
(0.14, 0.0177538)
(0.15, 0.0170905)
(0.16, 0.0164913)
(0.17, 0.0159469)
(0.18, 0.0154501)
(0.19, 0.0149948)
(0.2, 0.014576)
(0.21, 0.0141898)
(0.22, 0.0138327)
(0.23, 0.0135017)
(0.24, 0.0131945)
(0.25, 0.012909)
(0.26, 0.0126432)
(0.27, 0.0123958)
(0.28, 0.0121653)
(0.29, 0.0119506)
(0.3, 0.0117506)
(0.31, 0.0115644)
(0.32, 0.0113913)
(0.33, 0.0112305)
(0.34, 0.0110815)
(0.35, 0.0109437)
(0.36, 0.0108166)
(0.37, 0.0106998)
(0.38, 0.0105929)
(0.39, 0.0104956)
(0.4, 0.0104076)
(0.41, 0.0103287)
(0.42, 0.0102586)
(0.43, 0.0101972)
(0.44, 0.0101443)
(0.45, 0.0100997)
(0.46, 0.0100634)
(0.47, 0.0100353)
(0.48, 0.0100152)
(0.49, 0.0100032)
(0.5, 0.00999917)
(0.51, 0.0100031)
(0.52, 0.0100151)
(0.53, 0.0100352)
(0.54, 0.0100633)
(0.55, 0.0100995)
(0.56, 0.0101441)
(0.57, 0.0101969)
(0.58, 0.0102583)
(0.59, 0.0103283)
(0.6, 0.0104072)
(0.61, 0.0104951)
(0.62, 0.0105923)
(0.63, 0.0106992)
(0.64, 0.0108159)
(0.65, 0.010943)
(0.66, 0.0110807)
(0.67, 0.0112297)
(0.68, 0.0113903)
(0.69, 0.0115634)
(0.7, 0.0117494)
(0.71, 0.0119493)
(0.72, 0.0121639)
(0.73, 0.0123943)
(0.74, 0.0126416)
(0.75, 0.0129072)
(0.76, 0.0131925)
(0.77, 0.0134995)
(0.78, 0.0138303)
(0.79, 0.0141871)
(0.8, 0.0145731)
(0.81, 0.0149914)
(0.82, 0.0154463)
(0.83, 0.0159427)
(0.84, 0.0164865)
(0.85, 0.017085)
(0.86, 0.0177476)
(0.87, 0.0184857)
(0.88, 0.0193143)
(0.89, 0.0202529)
(0.9, 0.0213277)
(0.91, 0.0225749)
(0.92, 0.0240458)
(0.93, 0.0258166)
(0.94, 0.0280059)
(0.95, 0.0308116)
(0.96, 0.0345943)
(0.97, 0.0401049)
(0.98, 0.0492762)
(0.99, 0.0696867)
(0.991, 0.0734127)
(0.992, 0.0777989)
(0.993, 0.0830672)
(0.994, 0.0895601)
(0.995, 0.0978419)
(0.996, 0.108922)
(0.997, 0.124848)
(0.998, 0.150646)
(0.999, 0.204072)
(1, 0.5)
};

\legend{$\vri=2000$\\$\vri=200$\\$\vri=20$\\$\vri=2$\\$\vri=0.2$\\}
\end{axis}
\end{tikzpicture}
\caption{Dependence of the normalized group velocity of a one-dimensional periodic acoustic waveguide on the volume fraction $\delta$ of the scattering medium. The top solid black line corresponds to equal impedances of the two media. Every subsequent curve below corresponds to a tenfold decrease of the impedance of the scattering medium. The wave frequency for each curve is chosen in such a way to be in the middle of the first band for $\delta = 0.5$.
}
\label{graph2}
\end{figure}
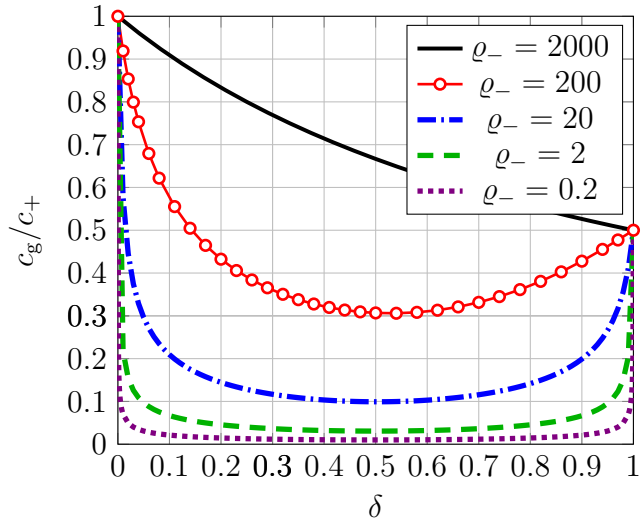

Graphs of the group velocities for different ratios of the impedances are shown in Figure \ref{graph2} where the top black solid curve corresponds to a periodic acoustic medium comprised of the scattering
acoustic material with $\cm = 1000$ m/s, $\vri = 2000$ kg/m$^3$ and the volume fraction $\delta$ in the host medium
with $\cm = 2000$ m/s and $\vri = 1000$ kg/m$^3$. The speed of propagation for matched impedances is given by \rf{cave}. Every subsequent curve below corresponds to a tenfold decrease of the density $\vri$ while other parameters are kept the same. To avoid confusion with band-edge effects, the frequencies of waves propagation were chosen in the center of the first band for $\delta = 0.5$ and are equal $1200$, $400$, $130$, and $40$ s$^{-1}$, respectively.
On the other hand, similar graphs for a fixed frequency $\om = 10$ s$^{-1}$ are practically indistinguishable. \\

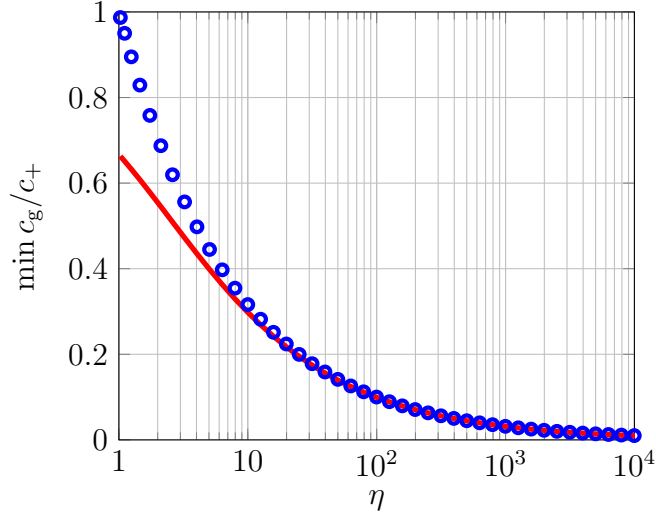
\begin{figure}[H]
\centering
\begin{tikzpicture}
\begin{axis}[
% scale only axis,
xmode=log,
log ticks with fixed point,
% width=0.4\textwidth,
xlabel=$\eta$,
xmin=1e-0, xmax=1e4,
ymin=0.0, ymax=1.0,
ylabel=$\min \cgr/\cp$,
xtick={1, 1e1, 1e2, 1e3, 1e4, 1e5},
xticklabels = {$1$,$10$,$10^2$,$10^3$,$10^4$,$10^5$},
grid=major,
grid=both,
xminorticks=true,
xmajorgrids,
xminorgrids,
]
% \addplot+[draw=red, line width= 2pt,only marks, mark=triangles] plot coordinates{
\addplot+[blue,ultra thick,only marks,mark=o,draw=blue,fill=blue] plot coordinates{
% approximation
(1.02663, 0.986947)
(1.10793, 0.950046)
(1.24822, 0.895063)
(1.455, 0.829028)
(1.73925, 0.758261)
(2.11613, 0.687431)
(2.6057, 0.619495)
(3.23403, 0.556068)
(4.03459, 0.497852)
(5.05, 0.444994)
(6.33434, 0.397328)
(7.95601, 0.354529)
(10.0014, 0.316206)
(12.5793, 0.281949)
(15.8272, 0.251361)
(19.9179, 0.224067)
(25.0693, 0.199723)
(31.5558, 0.178017)
(39.7227, 0.158665)
(50.005, 0.141414)
(62.9502, 0.126038)
(79.2478, 0.112333)
(99.7656, 0.100117)
(125.596, 0.0892301)
(158.115, 0.0795267)
(199.055, 0.0708784)
(250.595, 0.0631705)
(315.479, 0.0563008)
(397.165, 0.0501782)
(500, 0.0447213)
(629.463, 0.0398579)
(792.447, 0.0355234)
(997.631, 0.0316603)
(1255.94, 0.0282173)
(1581.14, 0.0251487)
(1990.54, 0.0224138)
(2505.94, 0.0199763)
(3154.79, 0.0178039)
(3971.64, 0.0158677)
(5000, 0.0141421)
(6294.63, 0.0126042)
(7924.47, 0.0112335)
(9976.31, 0.0100119)
(12559.4, 0.00892308)
(15811.4, 0.00795271)
(19905.4, 0.00708786)
(25059.4, 0.00631706)
(31547.9, 0.00563009)
(39716.4, 0.00501782)
(50000, 0.00447214)
};

\addplot+[draw=red, line width= 2pt,no marks, forget plot] plot coordinates{
% exact values
(1.02663, 0.66276)
(1.10793, 0.651232)
(1.24822, 0.632678)
(1.455, 0.608019)
(1.73925, 0.578386)
(2.11613, 0.545047)
(2.6057, 0.509268)
(3.23403, 0.472241)
(4.03459, 0.435001)
(5.05, 0.398406)
(6.33434, 0.363107)
(7.95601, 0.329575)
(10.0014, 0.298113)
(12.5793, 0.268895)
(15.8272, 0.241976)
(19.9179, 0.217337)
(25.0693, 0.194908)
(31.5558, 0.174576)
(39.7227, 0.156207)
(50.005, 0.139658)
(62.9502, 0.124781)
(79.2478, 0.11143)
(99.7656, 0.0994648)
(125.596, 0.0887543)
(158.115, 0.0791752)
(199.055, 0.0706132)
(250.595, 0.0629645)
(315.479, 0.0561347)
(397.165, 0.0500373)
(500, 0.0445959)
(629.463, 0.0397402)
(792.447, 0.0354076)
(997.631, 0.0315418)
(1255.94, 0.0280921)
(1581.14, 0.0250145)
(1990.54, 0.0222671)
(2505.94, 0.0198143)
(3154.79, 0.017624)
(3971.64, 0.0156672)
(5000, 0.0139177)
(6294.63, 0.0123526)
(7924.47, 0.010951)
(9976.31, 0.00969415)
(12559.4, 0.00856525)
(15811.4, 0.00754924)
(19905.4, 0.00663213)
(25059.4, 0.00580118)
(31547.9, 0.00504452)
(39716.4, 0.00435064)
(50000, 0.00370806)
};
\end{axis}
\end{tikzpicture}
\caption{Dependence of the normalized minimum group velocity $\min \cgr$ on the impedance contrast $\eta$ \rf{sigma}
when $\om = 10$s$^{-1}$ and $\delta = 0.5$ in a one-dimensional periodic waveguide. The solid red line corresponds to
the exact value $\cgr$ given by \rf{Cgr}. The blue circles represent the asymptotic formula \rf{cgr_min}.}
\label{asymptotics}
\end{figure}

We verify formula \rf{cgr_min} in Figure \ref{asymptotics}. The solid red line  corresponds to the exact value of the minimal group velocity
given by \rf{Cgr} while the blue circles represent the asymptotic formula \rf{cgr_min}. 
Below we show that formulas \rf{Cgr} and its approximation \rf{cgr_min} are valid for electromagnetic and elastic waves as well.

\section{Electromagnetic waves}

Propagation of electromagnetic waves through a one-dimensional periodic stacked medium
with period $\ell$ consisting of two types of alternating layers is also described by equations \rf{eq1}-\rf{eq2}, where
\begin{align}
 k_{\pm} = \frac{\om}{c}\,\sqrt{\ep_{\pm}\mu_{\pm}},
\end{align}
$c$ is the speed of light and $\ep_{\pm}, \mu_{\pm}$ are the relative permittivity and permeability of the layers, respectively.
For the magnetic component $u$ of the electromagnetic field %(see Fig. \ref{PC}) 
boundary conditions have the form
\begin{align}
% \label{bc}
\um = \up, \quad \frac{1}{\ep_{+}}\, \up_x  = \frac{1}{\ep_{-}}\, \um_x.
\end{align} 
Together with the Floquet condition \rf{eq4} we obtain the characteristic equation \rf{char_eqn}, where
\begin{align}
 z_{\pm} = \sqrt{\frac{\mu_{\pm}}{\ep_{\pm}}} 
\end{align}
is the wave impedance of the corresponding medium.
From \rf{cs} we obtain the group velocity
\begin{align}
 \cgr = \cp  \left(1-\frac{1}{2}\, \left( \frac{\mum}{\mup} + \frac{\epm}{\epp}\right) \delta + O\left(\delta^2 \right)\right), \quad \delta \ll 1,
 \label{cs1}
\end{align}
where $\cp = \dfrac{c}{\sqrt{\epp \mup}}$. Calculation of the minimum group velocity from \rf{cgr_min} yields
\begin{align}
 \min \cgr = \frac{c}{\sqrt{\epm \mup + \epp \mum}}.
\end{align}

\section{Elastic waves}

One-dimensional time-harmonic propagation of elastic waves through a periodic in the $x$-direction stack of layers  is governed by the equations
\cite{Landafshits7}
\begin{align}
\label{eq5}
 \frac{\D^2 u_1}{\D x^2} + \frac{\omega^2}{c^2_{p}}\, u_1 &= 0, \\[2mm]
 \frac{\D^2 u_\alpha}{\D x^2} + \frac{\omega^2}{c^2_{s}}\,  u_\alpha &= 0,\quad \alpha = 2,3,
 \label{eq6}
\end{align}
where the displacement vector $\u = (u_1, u_2, u_3)$ depend only on $x$ and
\begin{align}
 c_p = \sqrt{\frac{\lambda + 2\mu}{\rho}}, \quad c_s = \sqrt{\frac{\mu}{\rho}}
\end{align}
are velocities of the dilatational and shear waves. In this section $\mu$ and $\lambda$ denote the Lam\'{e} parameters and $\rho$ is the mass density. In addition, the displacement vector $\u$ and the normal component of the stress tensor $\bsigma$ must be continuous across the interfaces
\begin{align}
 \left. \left\llbracket \u \right\rrbracket \right. =\z, \quad
 \left. \left\llbracket \bsigma \cdot \n \right\rrbracket \right. =\z.
\end{align}
The latter condition in terms of the displacement components reads
\begin{align}
\label{bc1}
 (\lambda_{-} + 2\mu_{-}) \frac{\D u_{1}^{-}}{\D x} &= (\lambda_{+} + 2\mu_{+}) \frac{\D u_{1}^{+}}{\D x}, \\[2mm]
 \mu_{-} \frac{\D u_{\alpha}^{-}}{\D x} &= \mu_{+} \frac{\D u_{\alpha}^{+}}{\D x}, 
 \quad \alpha = 2,3.
 \label{bc2}
\end{align}
As a result, we obtain the same characteristic equation as in \rf{char_eqn} with the impedance
$z^{\pm}_p = \rho_{\pm} c_{p}^\pm$ for dilatational waves and $z^{\pm}_s = \rho_{\pm} c_{s}^\pm$ 
for shear waves. Formulas \rf{Cgr}, \rf{cs}-\rf{cgr_min} are valid for the two types of waves.

\section{Discussion}

The one-dimensional model considered above agrees with the numerical results obtained in two and three-dimensional cases \cite{Kafesaki:00,Krokhin:2003,Torrent:2006,Torrent:2007,GV:2019}.
Unlike a slowing down around inflection points of the dispersion curve in \cite{FV:06}, in our case a slowing down
happens for a wide frequency range.
Since the leading terms of the asymptotics do not depend on the frequency, the slowdown of the group velocity is not related to the resonant phenomenon \cite{Kafesaki:00} which occurs only at specific frequencies, but rather to a strong scattering and destructive interference of the waves.


\begin{thebibliography}{10}

\bibitem{Ruffa:92}
A.~A. Ruffa,
\newblock The Journal of the Acoustical Society of America {\bf 91}, 1 (1992).

\bibitem{Kafesaki:00}
M.~Kafesaki, R.~S. Penciu, and E.~N. Economou,
\newblock Phys. Rev. Lett. {\bf 84}, 6050 (2000).

\bibitem{Krokhin:2003}
A.~A. Krokhin, J.~Arriaga, and L.~N. Gumen,
\newblock Phys. Rev. Lett. {\bf 91}, 264302 (2003).

\bibitem{Torrent:2006}
D.~Torrent and J.~S\'anchez-Dehesa,
\newblock Phys. Rev. B {\bf 74}, 224305 (2006).

\bibitem{Torrent:2007}
D.~Torrent and J.~S{\'{a}}nchez-Dehesa,
\newblock New Journal of Physics {\bf 9}, 323 (2007).

\bibitem{Leroy:2009}
V.~Leroy et~al.,
\newblock {Applied Physics Letters} {\bf {95}} ({2009}).

\bibitem{Skvortsov:2019}
A.~Skvortsov, I.~MacGillivray, G.~S. Sharma, and N.~Kessissoglou,
\newblock Physical Review E {\bf 99} (2019).

\bibitem{Belyaev:87}
A.~Y. Belyaev,
\newblock Soviet Doklady {\bf 296}, 828 (1987),
\newblock (in Russian).

\bibitem{Ye:03}
Z.~Ye,
\newblock Acta Acustics united with Acustica {\bf 89}, 435 (2003).

\bibitem{Torrent:2012}
E.~Reyes-Ayona, D.~Torrent, and J.~S{\'{a}}nchez-Dehesa,
\newblock The Journal of the Acoustical Society of America {\bf 132}, 2896
  (2012).

\bibitem{Boutin:2013}
C.~Boutin,
\newblock Journal of the Acoustical Society of America {\bf 134}, 4717 (2013).

\bibitem{Humphrey:1998}
P.~A. Chinnery and V.~F. Humphrey,
\newblock The Journal of the Acoustical Society of America {\bf 103}, 1296
  (1998).

\bibitem{Wu:02}
F.~Wu, Z.~Liu, and Y.~Liu,
\newblock Phys. Rev. E {\bf 66}, 046628 (2002).

\bibitem{Bennetts:2019}
L.~G. Bennetts, M.~A. Peter, and R.~V. Craster,
\newblock Philosophical Transactions of the Royal Society A: Mathematical,
  Physical and Engineering Sciences {\bf 377}, 20190104 (2019).

\bibitem{Hale:80}
J.~K. Hale,
\newblock {\em Ordinary Differential Equations},
\newblock John Wiley \& Sons, Inc., 1980.

\bibitem{Ye:00}
Z.~Ye and E.~Hoskinson,
\newblock Applied Physics Letters {\bf 77}, 4428 (2000).

\bibitem{Vasseur:01}
J.~O. Vasseur et~al.,
\newblock Phys. Rev. Lett. {\bf 86}, 3012 (2001).

\bibitem{Khelif:04}
A.~Khelif, A.~Choujaa, S.~Benchabane, B.~Djafari-Rouhani, and V.~Laude,
\newblock Applied Physics Letters {\bf 84}, 4400 (2004).

\bibitem{Olsson:09}
R.~H. Olsson, III and I.~El-Kady,
\newblock {Measurement Science and Technology} {\bf {20}}, 012002 ({2009}).

\bibitem{Cocoletzi:94}
R.~Esquivel-Sirvent and G.~H. Cocoletzi,
\newblock The Journal of the Acoustical Society of America {\bf 95}, 86 (1994).

\bibitem{Sigalas:95}
M.~M. Sigalas and C.~M. Soukoulis,
\newblock Phys. Rev. B {\bf 51}, 2780 (1995).

\bibitem{Hussein:06}
M.~I. Hussein, G.~M. Hulbert, and R.~A. Scott,
\newblock Journal of Sound and Vibration {\bf 289}, 779  (2006).

\bibitem{Boudouti:09}
E.~E. Boudouti, B.~Djafari-Rouhani, A.~Akjouj, and L.~Dobrzynski,
\newblock Surface Science Reports {\bf 64}, 471  (2009).

\bibitem{Bendickson:96}
J.~M. Bendickson, J.~P. Dowling, and M.~Scalora,
\newblock Phys. Rev. E {\bf 53}, 4107 (1996).

\bibitem{Yeh:05}
P.~Yeh,
\newblock {\em Optical waves in layered media},
\newblock Wiley, 2005.

\bibitem{Bradley:94}
C.~E. Bradley,
\newblock The Journal of the Acoustical Society of America {\bf 96}, 1844
  (1994).

\bibitem{Adams:08}
S.~D. Adams, R.~V. Craster, and S.~Guenneau,
\newblock Proceedings of the Royal Society A: Mathematical, Physical and
  Engineering Sciences {\bf 464}, 2669  (2008).

\bibitem{GV:2019}
Y.~A. Godin and B.~Vainberg,
\newblock Dispersion of waves in two and three-dimensional periodic media,
\newblock eprint arXiv:1905.12529, 2019.

\bibitem{YG:06}
Y.~A. Godin,
\newblock Waves in Random and Complex Media {\bf 16}, 409 (2006).

\bibitem{Landafshits7}
L.~D. Landau and E.~M. Lifshit︠s︡,
\newblock {\em Theory of Elasticity}, volume~7 of {\em Course of theoretical
  physics},
\newblock Pergamon Press, Oxford, 2d english ed., rev. and enl. edition, 1970.

\bibitem{FV:06}
A.~Figotin and I.~Vitebskiy,
\newblock Waves in Random and Complex Media {\bf 16}, 293 (2006).

\end{thebibliography}
\end{document}